\documentclass{aa}

\usepackage{siunitx}
\usepackage{caption}
\usepackage{multicol}
\DeclareSIUnit{\angstrom}{\textup{\AA}}
\usepackage{bm}
\usepackage{ulem}

\usepackage{graphicx}
\usepackage[varg]{txfonts}

\begin{document} 

   \title{Estimating the lateral speed of a fast shock driven by a coronal mass ejection at the location of solar radio emissions}
   \titlerunning{Estimating the lateral speed of a fast shock driven by a coronal mass ejection}

   \author{S. Normo\inst{1,2}
          \and
          D. E. Morosan\inst{1,2}
          \and
          E. K. J. Kilpua\inst{1}
                    \and
          J. Pomoell \inst{1}
          }

   \institute{Department of Physics, University of Helsinki, P.O. Box 64, FI-00014 Helsinki, Finland \\
              \email{diana.morosan@helsinki.fi}
            \and
             Department of Physics and Astronomy, University of Turku, 20014, Turku, Finland
              }

   \date{Received  }

  \abstract
   { Fast coronal mass ejections (CMEs) can drive shock waves capable of accelerating electrons to high energies. These shock-accelerated electrons act as sources of electromagnetic radiation, often in the form of solar radio bursts. Recent findings suggest that radio imaging of solar radio bursts can provide a means to estimate the lateral expansion of CMEs and associated shocks in the low corona. }
   { Our aim is to estimate the expansion speed of a CME-driven shock at the locations of radio emission using 3D reconstructions of the shock wave from multiple viewpoints.}
   { In this study, we estimated the 3D location of radio emission using radio imaging from the Nançay Radioheliograph and the 3D location of a CME-driven shock. The 3D shock was reconstructed using white-light and extreme ultraviolet images of the CME from the Solar Terrestrial Relations Observatory, Solar Dynamics Observatory, and the Solar and Heliospheric Observatory. The lateral expansion speed of the CME-driven shock at the electron acceleration locations was then estimated using the approximate 3D locations of the radio emission on the surface of the shock.}
   {The radio bursts associated with the CME were found to reside at the flank of the expanding CME-driven shock. We identified two prominent radio sources at two different locations and found that the lateral speed of the shock was between $\SI{800}{}$ and $\SI{1000}{\kilo\meter\per\second}$ at these locations. Such a high speed during the early stages of the eruption already indicates the presence of a fast shock in the low corona. We also found a larger ratio between the radial and lateral expansion speed compared to values obtained higher up in the corona.} 
   {We estimated for the first time the 3D expansion speed of a CME-driven shock at the location of the accompanying radio emission. The high shock speed obtained is indicative of a fast acceleration during the initial stage of the eruption. This acceleration leading to lateral speeds in the range of $\SI{800}{}-\SI{1000}{\kilo\meter\per\second}$ is most likely one of the key parameters contributing to the presence of metric radio emissions, such as type II radio bursts.}


   \maketitle

\section{Introduction}

Coronal mass ejections (CMEs) are one of the most energetic phenomena in the Solar System \citep[e.g.][]{webb2012}. They often drive coronal shock waves that can in turn accelerate particles up to relativistic energies \citep[e.g.][]{reames1999}. The fast lateral expansion of CMEs and associated shock waves has recently been reported to influence the occurrence of solar radio bursts associated with CMEs \citep[e.g.][]{morosan2021}.

Solar radio bursts are transient increases in the solar radio emission intensity above the background level. They can be categorised based on their properties and appearance in a dynamic spectrum into types I to V \citep{Wild1963}. In particular, type II and type IV radio bursts are most commonly associated with CMEs \citep[e.g.][]{Gergely1986,Classen2002,Kumari_2021,morosan2021,Kumari2023}. In dynamic spectra, type II radio bursts appear as lanes drifting slowly (drift rate typically less than $\SI{1}{\mega\hertz\per\second}$) towards lower frequencies \citep{ro59,ne85,Cairns2003}. There are often two bands of emission with a frequency ratio of approximately two representing emission at the local plasma frequency and its harmonic. Type II bursts are closely associated with CME-driven shocks that accelerate electrons to high energies \citep{ma05,zu18}. In addition, approximately $20\%$ of type II bursts exhibit fine structures called herringbones \citep{ro59,holman1983,ca87}. They are characterised by a much faster drift to lower and/or higher frequencies superimposed on a type II burst or even occurring on their own. Herringbone bursts originate from electron beams accelerated at the flanks of CME-driven shocks as they expand in the solar corona \citep[e.g.][]{ca13,mo19a,morosan2022,Zhang2024}. In contrast, type IV radio bursts are represented by a longer duration broadband continuum emission in dynamic spectra, and they can be either moving or stationary in nature, both of which have been associated with CMEs \citep{Weiss1963,Pick1986}. While plasma emission is a widely accepted emission mechanisms for type II bursts \citep[e.g.][]{Melrose1980}, the formation of type IV radio emission is less certain. Several different emission mechanisms have been reported to be responsible for type IV bursts, including synchrotron and gyrosynchrotron \citep{Boischot1968,Dulk1973,tu13,ca17}, in addition to plasma emission \citep{Benz1976,Gary1985,mo19b,Morosan2020b,salas2020polarisation}. 

\begin{figure*}[h!]
    \centering
    \includegraphics[width=0.85\textwidth]{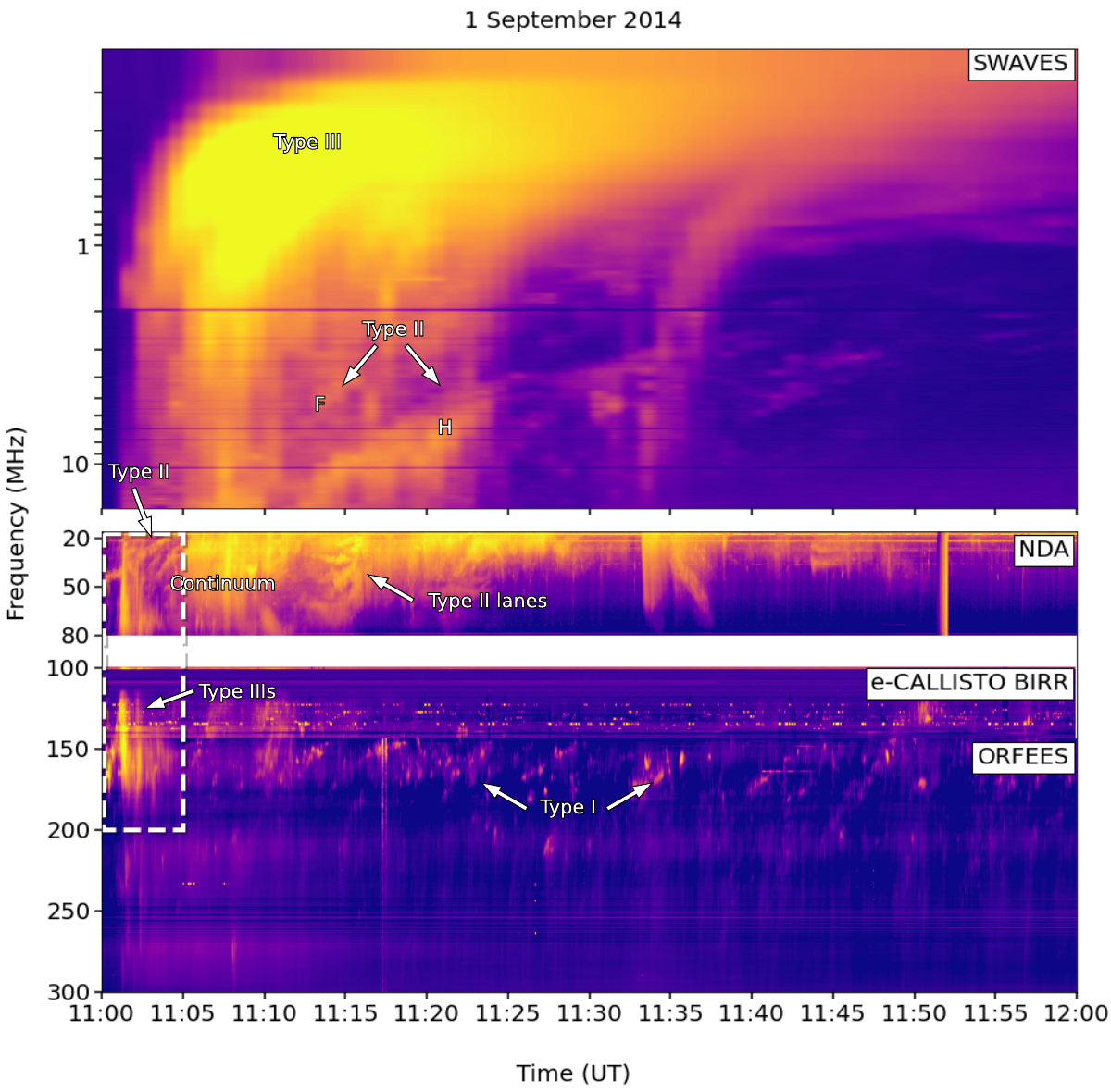}
    \caption{Dynamic spectrum from STEREO-B/SWAVES, NDA, e-CALLISTO Birr, and ORFEES between \SI{0.125}{\mega\hertz} and \SI{300}{\mega\hertz}. The spectrum shows various types of solar radio bursts labelled and indicated with arrows. The dashed rectangle indicates the part of the full spectrum shown in Fig. \ref{fig:zoom-in_ds_and_contours}a.}
    \label{fig:full_ds}
\end{figure*}

\begin{figure*}[h!]
    \centering
    \includegraphics[width=0.85\textwidth]{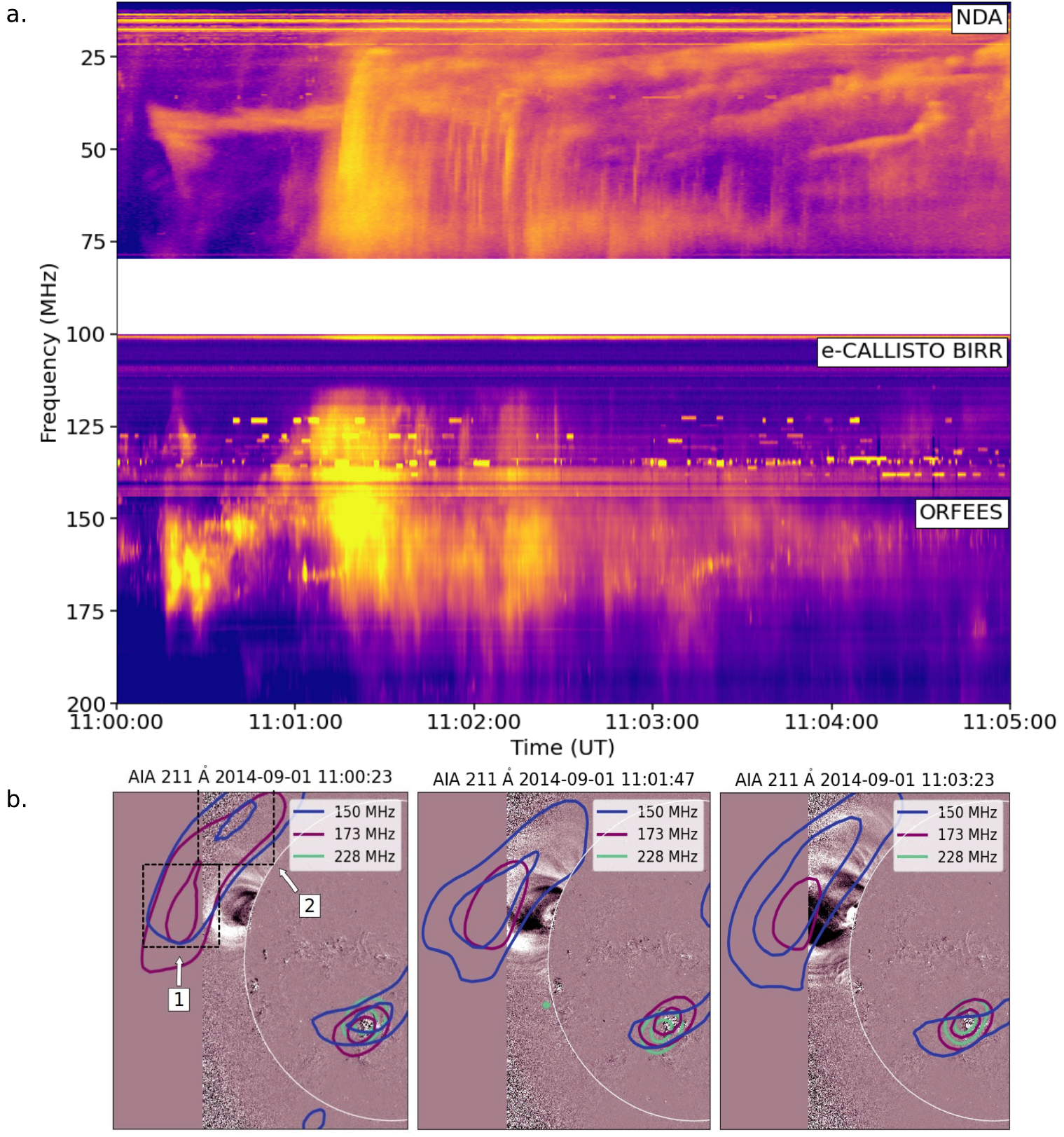}
    \caption{Dynamic spectrum and contours of the herringbone-like radio emission. (a) Zoomed-in dynamic spectrum from NDA, e-CALLISTO Birr, and ORFEES between \SI{10}{\mega\hertz} and \SI{200}{\mega\hertz}. {This part of the spectrum shows the herringbone-like fine structure observed by ORFEES as well as the first type II burst observed by NDA.} (b) Contours of the radio emission at $30\,\%$ and $70\,\%$ levels from NRH Stokes $I$ (total intensity) observations at $\SI{150}{\mega\hertz}$ (blue), $\SI{173}{\mega\hertz}$ (purple), and $\SI{228}{\mega\hertz}$ (green). The contours are overlaid on running difference images observed by SDO/AIA at $\SI{211}{\angstrom}$. The time interval between the images used to produce the running difference images is $\SI{2}{\min}$. The two peaks in radio intensity in the left panel of (c) are outlined by rectangles and coined Box 1 and Box 2.} 
    \label{fig:zoom-in_ds_and_contours}
\end{figure*}

The measured speed of a CME is a combination of its radial speed (the speed of the nose of the CME) and lateral expansion speed \citep[e.g.][]{2003DalLago,Schwenn2005,Gopalswamy2009expansion,Dagnew2020}. Typically, the radial and lateral expansion speeds are estimated using white-light images. However, a statistical study \citep{morosan2021} as well as studies of individual events \citep[e.g.][]{de12,mo19a} suggest that the lateral expansion of a CME or the accompanying shock wave in the low corona could be estimated using radio emission. Statistical studies have also shown that wider CMEs are more likely to be associated with metric radio emission \citep[e.g.][]{Lara2003,Kahler2019,Kumari2023} than narrow CMEs, indicating that a fast expansion of the CME in the low corona could be crucial for the formation of a shock wave in the low corona and for subsequent electron acceleration to occur. However, the lateral expansion speeds of CME-driven shocks in the low corona have been poorly studied and even less so in 3D due to the limited number of multi-point imaging observations where the shock outline is clear.

In this paper, we present the analysis of herringbone-like radio emission associated with a CME that erupted on 1 September 2014. The CME drove a shock wave that showed a clear boundary in multi-vantage point extreme ultraviolet (EUV) imaging and allowed the shock to be fitted in 3D. Using a 3D reconstruction of the CME-driven shock, we found the radio emission originates from the flank of the expanding shock. We then estimated the lateral expansion speed of the shock at the location of said radio emission.   

\section{Observations and data analysis}

\begin{figure}[h!]
    \centering
    \includegraphics[width=\linewidth]{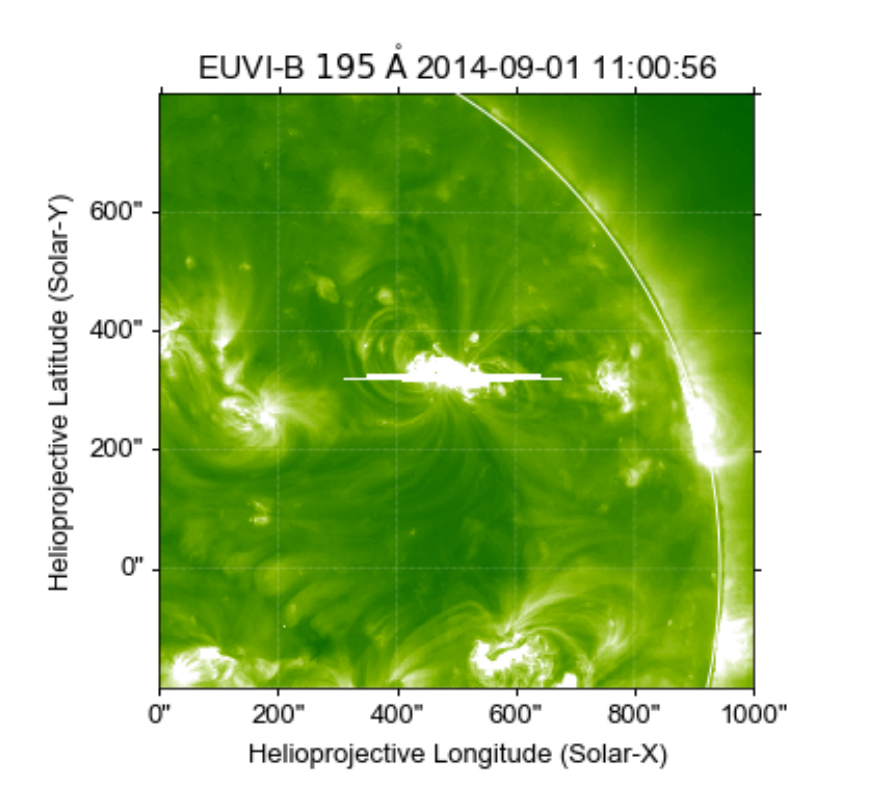}
    \caption{Active region where the eruption on 1 September 2014 originated and an accompanying flare observed by STEREO/SECCHI/EUVI-B at a wavelength of $\SI{195}{\angstrom}$ at 11:00:56~UT.}
    \label{fig:STB_zoom-in}
\end{figure}

\subsection{Radio emission}

On 1 September 2014, a radio event was observed simultaneously with a CME. The start of the radio emission was at $\sim$11:00~UT, which corresponded to the early evolution of the CME. The associated radio emission is shown in the combined dynamic spectrum in Fig. \ref{fig:full_ds} observed by the Radio and Plasma Wave Investigation \citep[SWAVES;][]{Bougeret2008} instrument on board the Solar Terrestrial Relations Observatory Behind \citep[STEREO-B;][]{ka08}, the Nançay Decameter Array \citep[NDA;][]{bo80}, ORFEES (Observation Radio pour FEDOME et l’Étude des Éruptions Solaires) radio spectrograph \citep{orfees21}, and a radio spectrometer from the e-CALLISTO network \citep{Benz2005,Benz2009}: e-CALLISTO Birr. The spectrum covers a frequency range of $0.125-\SI{300}{\mega\hertz}$ between 11:00 and 12:00~UT. The instruments NDA, e-CALLSITO Birr, and ORFEES observed multiple type III bursts shortly after the start of the radio event. Type III bursts were also observed by STEREO-B/SWAVES. Additionally, at low frequencies ($\lesssim\SI{100}{\mega\hertz}$), both continuum-like emission and multiple type II features were observed. Notably, NDA observed a multi-lane type II burst between 11:00 and 11:05~UT, a continuum-like emission right after 11:05~UT, and type II lanes around 11:15~UT, whereas STEREO-B/SWAVES observed a type II burst shortly after 11:10~UT. The fundamental and harmonic lanes of this type II are marked in Fig. \ref{fig:full_ds} with F and H, respectively. A type I noise storm was present throughout the observations and most prominently seen at a frequency range of $\SI{144}{}-\SI{200}{\mega\hertz}$. Within the same frequency range, many fast-drifting individual bursts resembling herringbones were observed between 11:00 and 11:05~UT. As outlined by the dashed rectangle in Fig. \ref{fig:full_ds}, this part of the dynamic spectrum is detailed in Fig. \ref{fig:zoom-in_ds_and_contours} together with the e-CALLISTO Birr and NDA spectra. The type I noise storm and the type III bursts make it difficult to isolate and unambiguously distinguish the features that belong to the emission. 

In order to study the location of the observed herringbone-like emission relative to the eruption, interferometric radio observations from the Nançay Radioheliograph \citep[NRH;][]{ke97} were used to construct images of the radio bursts. However, the radio imaging was limited by the NRH observing frequencies to $\SI{150}{}-\SI{450}{\mega\hertz}$. Consequently, the NRH data was extracted as contours outlining the bright emission at $\SI{150}{\mega\hertz}$ (blue), $\SI{173}{\mega\hertz}$ (purple), and $\SI{228}{\mega\hertz}$ (green), as shown in {Fig. \ref{fig:zoom-in_ds_and_contours}b}. The contours are overlaid on running difference images observed by the Atmospheric Imaging Assembly \citep[AIA;][]{le12} on board the Solar Dynamics Observatory \citep[SDO;][]{pe12} at $\SI{211}{\angstrom}$ between 11:00 and 11:03 UT. The radio contours were drawn where the brightness temperature of the radio emission is at $30\,\%$ and $70\,\%$ of its maximum value.

\begin{figure}[h!]
   \centering
   \includegraphics[width=0.9\linewidth]{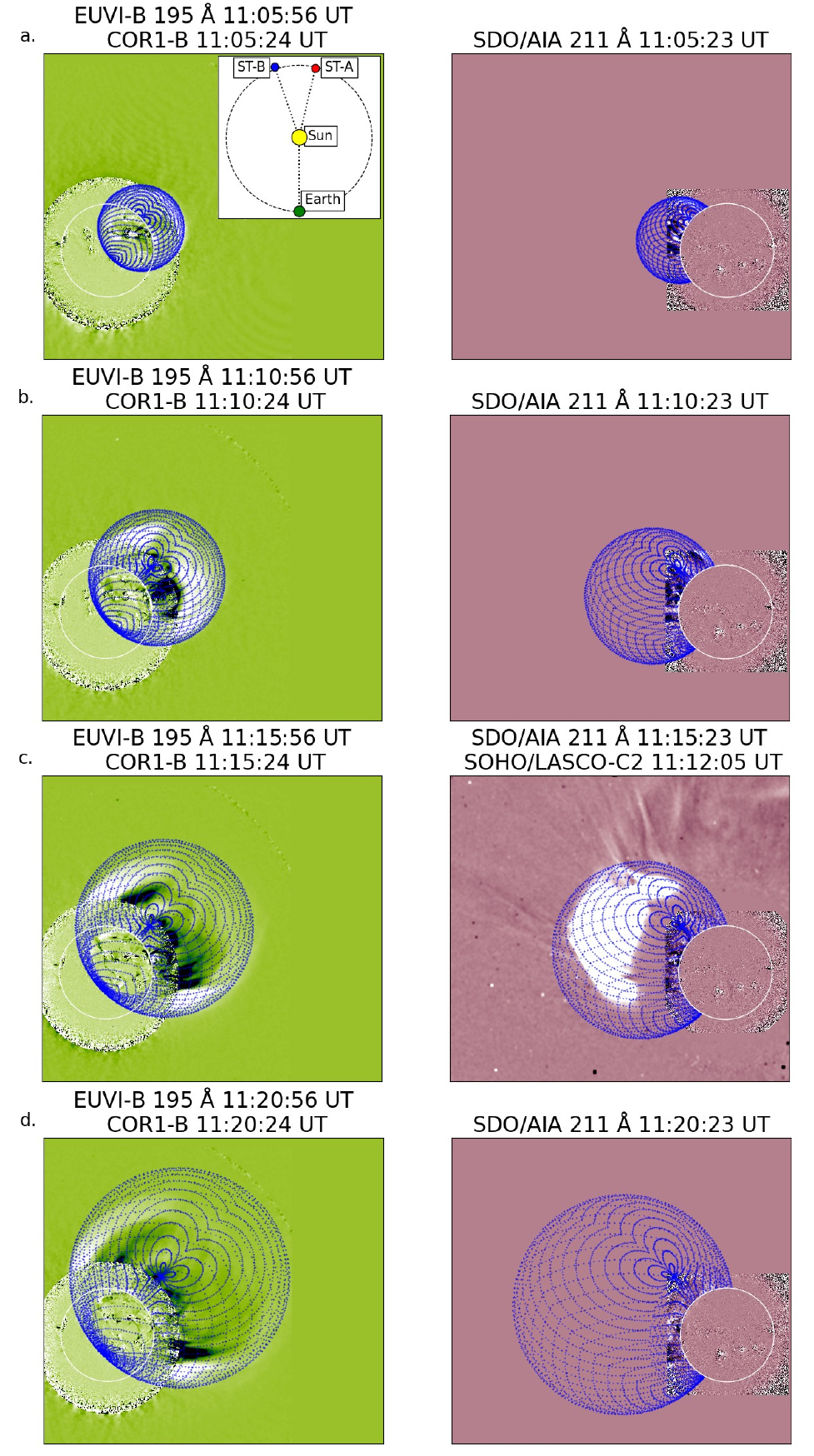}
   \caption{Fitting of the CME shock with the GCS model at four different times: (a) 11:05, (b) 11:10, (c) 11:15, and (d) 11:20 UT. The panels show the fit as seen from STEREO-B (left) and Earth (right). In addition, panel (a) shows the location of the observing spacecraft relative to the Sun. The GCS model (blue) is overlaid on the EUV and white-light running difference images. The EUV observations consist of STEREO/SECCHI/EUVI-B at $\SI{195}{\angstrom}$ and SDO/AIA at $\SI{211}{\angstrom}$. The white-light observations were mainly made by STEREO/SECCHI/COR1-B, with the exception of the right panel in (c) where SOHO/LASCO C2 data was available. The time interval between the images used to produce the running difference images is $\SI{5}{\min}$ for both EUVI-B and COR1-B and $\SI{12}{\min}$ for LASCO C2.}
   \label{fig:GCS_model}
 \end{figure}

 \begin{figure*}[h!]
   \centering
   \includegraphics[width=0.75\textwidth]{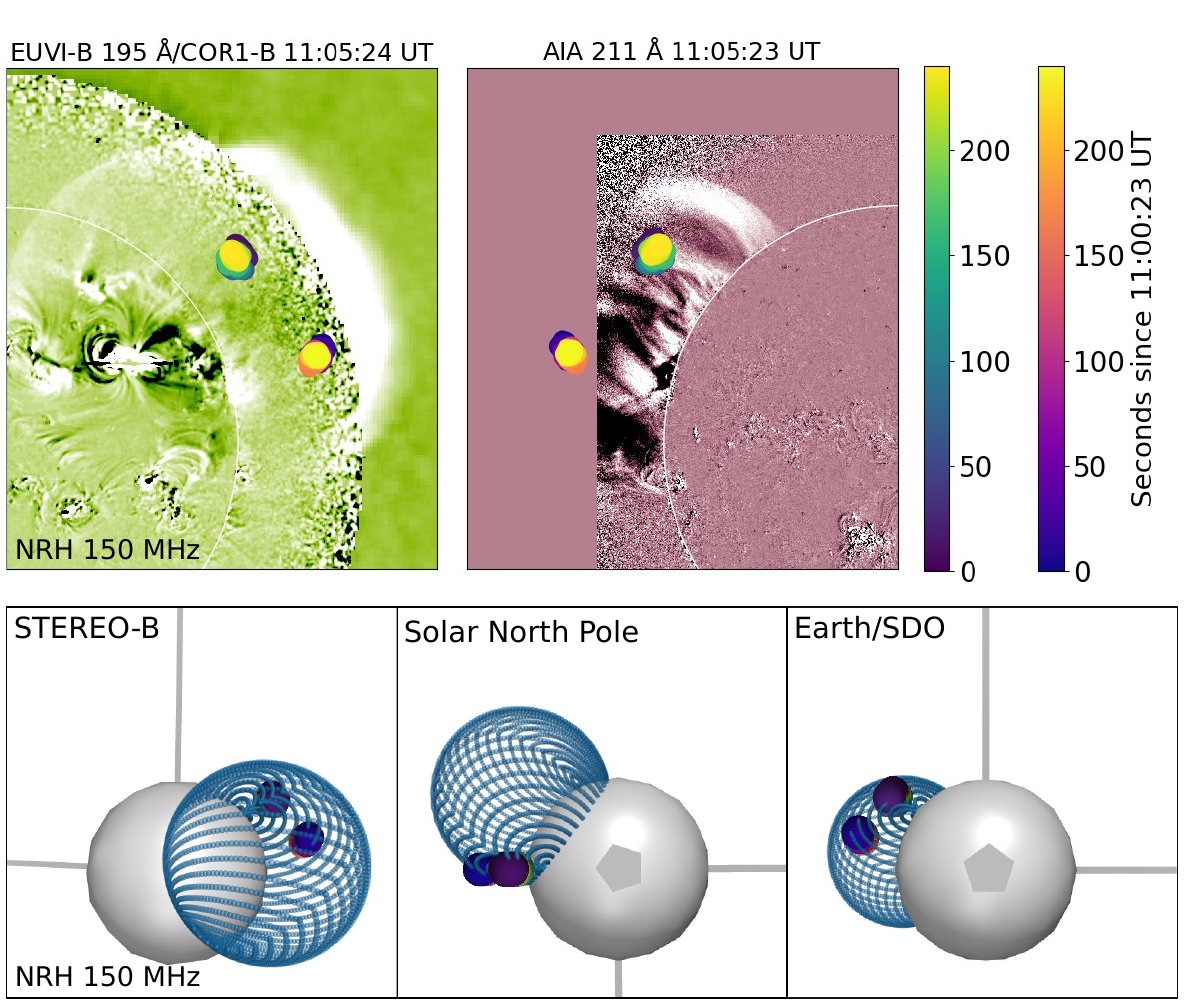}
   \caption{Centroids of the radio emission at $\SI{150}{\mega\hertz}$. The top panel shows the centroids of the radio emission from the perspective of STEREO-B (left) and Earth (right). The centroids are plotted over running difference images from STEREO/SECCHI/EUVI-B at $\SI{195}{\angstrom}$, STEREO/SECCHI/COR1-B, and SDO/AIA at $\SI{211}{\angstrom}$. The bottom panel shows the centroids together with the 3D reconstruction of the shock from three perspectives: STEREO-B (left), solar north pole (middle), and Earth (right).}
   \label{fig:centroids}
 \end{figure*}

\subsection{The coronal mass ejection}

The CME accompanying the radio event erupted behind the eastern limb of the Sun. Figure \ref{fig:STB_zoom-in} shows the active region and flare associated with the eruption in EUV at 11:00:56~UT observed by the Extreme UltraViolet Imager \citep[EUVI;][]{Wuelser2004} on board STEREO-B. In addition to SDO {and STEREO-B}, the CME was also observed by the Solar and Heliospheric Observatory \citep[SOHO;][]{do95}. In white-light, the CME was first observed at 11:05 UT by the Inner Coronagraph COR1 \citep{ho08} on board STEREO-B. A few minutes later, at 11:12 UT, the CME was observed in white-light from another viewpoint by the Large Angle and Spectrometric Coronagraph \citep[LASCO C2;][]{br95} on board SOHO. There were no observations from STEREO-A during this time period. The locations of the spacecraft relative to the Sun are shown in the left panel of Fig. \ref{fig:GCS_model}a. Additionally, the earlier stages of the eruption were imaged in EUV by STEREO-B's EUVI and SDO's AIA instruments. A prominent shock wave was seen in the white-light images. An estimate for the plane-of-sky speed of the CME from the SOHO/LASCO CME catalogue \citep{Gopalswamy2009} is $\SI{1901}{\kilo\meter\per\second}$ for a linear fit. Using a second-degree fit, a maximum speed of $\sim\SI{2600}{\kilo\meter\per\second}$ was found at a heliocentric distance of $\sim4\,R_\odot$.

\section{Results}

\subsection{Characteristics of the radio emission}

The zoomed-in ORFEES dynamic spectrum in Fig. \ref{fig:zoom-in_ds_and_contours}a shows herringbone-like radio emission observed during the early evolution of the CME. More precisely, we identified reverse (positive) drifting structures with a drift rate of $\mathrm{d}f/\mathrm{d}t \approx \SI{25.5}{\mega\hertz\per\second}$ (see Fig. \ref{fig:tracking_appendix}a in Appendix \ref{app:B}). The radio contours in Fig. \ref{fig:zoom-in_ds_and_contours}b reveal simultaneous bright emissions at $\SI{150}{\mega\hertz}$ (blue) and $\SI{173}{\mega\hertz}$ (purple) initially ahead of the CME and later overlapping the CME eruption as seen in the plane of sky. In addition, the left panel in Fig. \ref{fig:zoom-in_ds_and_contours}b shows two distinct bright peaks in the emission: one at $\SI{150}{\mega\hertz}$ and the other at $\SI{173}{\mega\hertz}$. The peaks were analysed separately, and they are outlined with dashed rectangles labelled as Box 1 and Box 2 in Fig.~\ref{fig:zoom-in_ds_and_contours}b. Persistent emission at all three frequencies ($\SI{150}{\mega\hertz}$, $\SI{173}{\mega\hertz}$, and $\SI{228}{\mega\hertz}$) were observed to overlap at an active region on the solar disc. This emission corresponds to the type I noise seen in the dynamic spectrum in Fig. \ref{fig:full_ds}.

\begin{figure}[h!]
   \includegraphics[width=\linewidth]{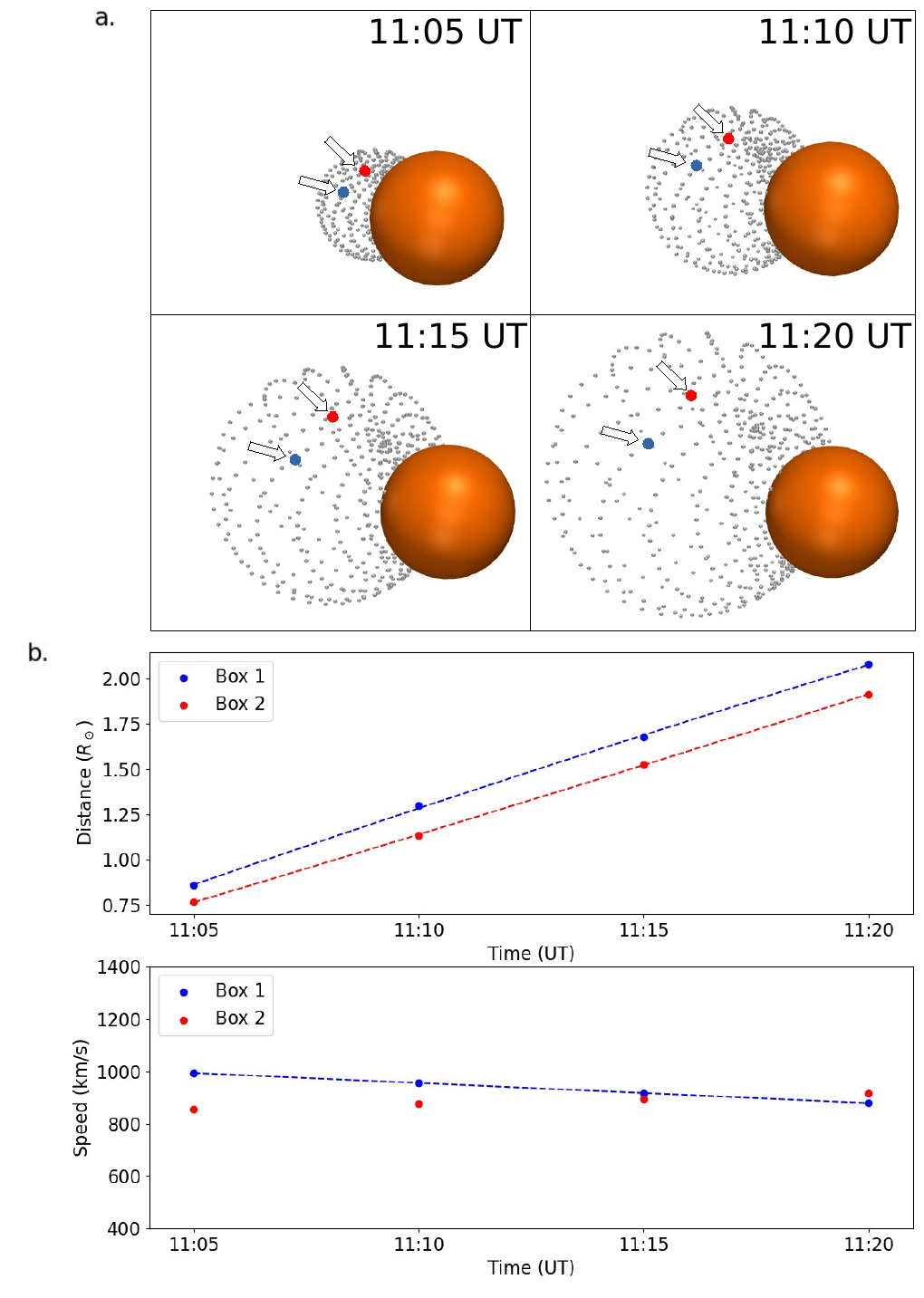}
   \caption{Lateral expansion of the shock at the locations of the radio sources. (a) Estimated locations of the radio emission on the 3D model of the shock at four different times: 11:05, 11:10, 11:15, and 11:20 UT. The coordinates of the points, which correspond to the radio emission peak intensities within Box 1 (blue) and Box 2 (red) and indicated by the white arrows, can be used to estimate the speed. (b) Distance from the shock centre line of the radio emission sources calculated from the coordinates of the points in (a). In the top panel, a second-degree polynomial fit is done to describe the movement. The bottom panel shows the speed calculated as a derivative of the fitted function in the top panel. The blue dashed line indicates a first-degree polynomial fit to the decelerating profile of the radio source in Box 1.}
   \label{fig:speed_estimate}
 \end{figure}

\subsection{Reconstruction of the CME-driven shock and location of the radio emissions in 3D} 

We reconstructed the CME shock in three dimensions using EUV and white-light images from two viewpoints and the PyThea tool \citep{Kouloumvakos2022}. The shock was fitted using the graduated cylindrical shell \citep[GCS;][]{the06,the09} model, which is typically used to fit CME flux ropes. We used a simplified version of this model by setting the half-angle to zero since the spheroid model typically used for fitting shocks was unable to simultaneously fit the features seen in both STEREO-B's and SDO's viewpoints. Figures \ref{fig:GCS_model}a-\ref{fig:GCS_model}d show the fitting of the CME shock at four different times and from two perspectives: STEREO-B (left) and SDO (right). The shock fittings (in blue) are plotted over running difference EUVI-B $\SI{195}{\angstrom}$ and COR1-B images in the left panels and over running difference SDO/AIA $\SI{211}{\angstrom}$ in the right panels. In addition, the right panel in Fig. \ref{fig:GCS_model}c includes a running difference SOHO/LASCO C2 image. 

Following the reconstruction of the CME-driven shock in 3D, we sought the spatio-temporal relation between the peak radio emission intensities located in Boxes 1 and 2 in Fig. \ref{fig:zoom-in_ds_and_contours}b and the shock surface. First, we determined the centroids of the observed bright emission at $\SI{150}{\mega\hertz}$ and $\SI{173}{\mega\hertz}$ from the NRH imaging. This provided the location of the radio emission in the plane of sky as viewed from Earth. In addition to Earth's view of the radio centroids, we also investigated how they are located from STEREO-B's perspective. In order to do a coordinate transform to the frame of reference of STEREO-B, 3D coordinates for each emission source were needed. The plane-of-sky $x$ and $y$ coordinates were obtained from NRH imaging, which means that the unknown $z$ coordinates needed to be estimated (see for example, the methods of \cite{Morosan2020b}). 

Figure \ref{fig:zoom-in_ds_and_contours}b shows that, initially, the location of the radio emission was outside the CME outline in EUV images, and as the eruption expanded, it overlapped the evolving CME in the plane of sky. Therefore, we could assume that the emission is outside of the eruption the entire time since the morphology of the radio emission did not change in the dynamic spectrum. We started by assuming that the radio bursts are emitted in the plane of sky (i.e.\ the \textit{z} coordinate of this emission is ${\sim}0$). However, we note that the radio sources could also move towards or away from the observer, from Earth's perspective, and only cross the $z=0$ plane as they propagate. It is unlikely that the centroid location is away from the plane of sky based on the following considerations: The z coordinate cannot be large and positive, as then the radio emission would be located too far away from the expanding eruption in 3D, and the $z$ coordinate also cannot be large and negative, as then it would be located inside the CME core, while initial observations place this emission outside of the CME eruption. Therefore, the assumption that the radio sources propagate in the plane of sky is a good approximation to determine their location in other perspectives and relative to the CME-driven shock. 

The top-right panel in Fig. \ref{fig:centroids} shows the centroids of the radio emission peak intensities located in Boxes 1 and 2 at $\SI{150}{\mega\hertz}$ for $\SI{4}{min}$ period starting at 11:00 UT. The centroids are overlaid on the SDO/AIA $\SI{211}{\angstrom}$ running difference image at 11:05 UT. From Earth's viewpoint, the centroids overlap with the eruption at this time and exhibit only very little apparent movement in time. The centroids at $\SI{173}{\mega\hertz}$ behave in a very similar manner (see Fig. \ref{fig:centroids_appendix} in Appendix \ref{app:A}). The left panel in Fig. \ref{fig:centroids} shows the coordinate transformed centroids at $\SI{150}{\mega\hertz}$ using $z\approx0$ from STEREO-B's point of view. The centroids are overlaid on the running difference EUVI-B $\SI{195}{\angstrom}$ and COR1-B images at 11:05 UT. The centroids of the radio emission in Boxes 1 and 2 are plotted with the reconstructed shock at $\SI{150}{\mega\hertz}$ in the bottom panel of Fig. \ref{fig:centroids}. The centroids and the shock are viewed from three perspectives: STEREO-B (left), solar north pole (middle), and Earth (right). The view from the north pole of the Sun shows that the centroids are located at the flank of the shock bubble.

 \begin{figure}
    \centering
    \includegraphics[width=\linewidth]{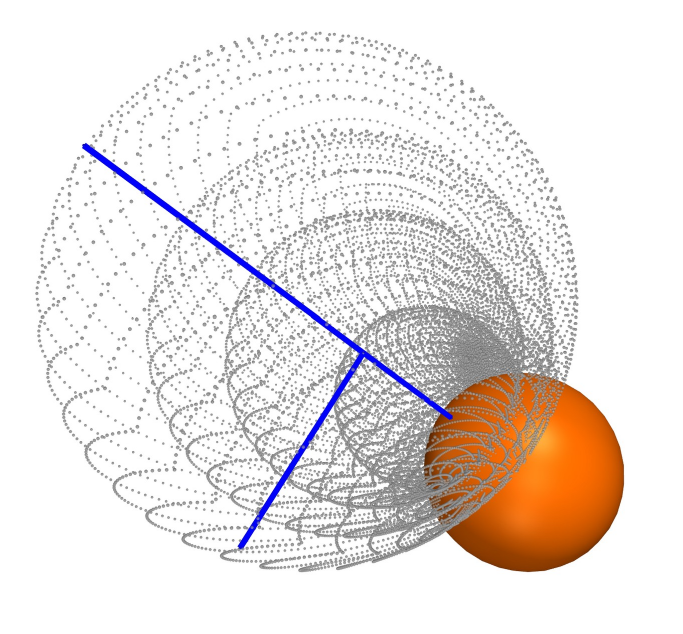}
    \caption{Three-dimensional models of the CME-driven shock at four different times (11:05, 11:10, 11:15, 11:20 UT) as viewed from the solar north pole. The blue line pointing away from the Sun is the centre line of the shock, which is used to estimate the radial speed of the shock. The blue line perpendicular to it shows an example of the distance of a point on the shock surface from the centre line. The shortest distance of each point in Fig. \ref{fig:speed_estimate}a is used to estimate the lateral expansion speed of the shock. Here, the distance between the location of the radio source in Box 1 at 11:20 UT and the centre line is shown.}
    \label{fig:radial_and_lateral_distance}
\end{figure}

\subsection{Estimating the lateral speed of the CME-driven shock} 

We found two distinct locations of radio emission at the flank of the CME-driven shock. The spatio-temporal co-location of the radio emission and the 3D reconstruction of the shock can be used to estimate the lateral expansions of the CME shock at the location of electron acceleration. Figure \ref{fig:speed_estimate}a shows a 3D plot of the of the reconstructed shock as viewed from Earth at four different times: 11:05 (top left), 11:10 (top right), 11:15 (bottom left), and 11:20~UT (bottom right). The points inside Box 1 (blue) and Box 2 (red) were determined on the surface of the shock at 11:05 UT. The location was estimated by finding the point in the shock mesh near the 3D coordinates of the centroids. These points were tracked for the following three shock reconstruction time steps and used in a distance-time plot to estimate the expansion speed.

We estimated the lateral expansion speed by first determining the centre line inside the shock, and then we calculated the distance of a point on the surface of the shock from the centre line. Figure \ref{fig:radial_and_lateral_distance} shows the shock reconstructions at the four different times plotted together with the shock centre line. The centre line was constructed using the coordinates of the shock model, and it points from the solar surface towards the apex of the shock. The distance from the centre line of each point in Fig. \ref{fig:speed_estimate}a was calculated in order to estimate the lateral expansion speed of the CME-driven shock. In Fig. \ref{fig:radial_and_lateral_distance}, the line perpendicular to the centre line shows an example distance of a point on the surface of the shock at 11:20 UT. 

The top panel in Fig. \ref{fig:speed_estimate}b shows the distances of the mesh points in Fig. \ref{fig:speed_estimate}a from the centre line as a function of time. The movement of these points was fitted with a second-degree polynomial function, and the fits are indicated by the dashed curves in the top panel of Fig. \ref{fig:speed_estimate}b. The bottom panel in Fig. \ref{fig:speed_estimate}b shows the derived speed of the shock at the location of the radio emissions. The speeds were obtained by calculating the derivatives of the fitted functions in the top panel at four different times corresponding to the times of the GCS shock models. At all time instances, the speed of the shock at the location of radio emissions in both Boxes 1 and 2 is between $\SI{800}{}$ and $\SI{1000}{\kilo\meter\per\second}$.

\begin{figure}
    \includegraphics[width=\linewidth]{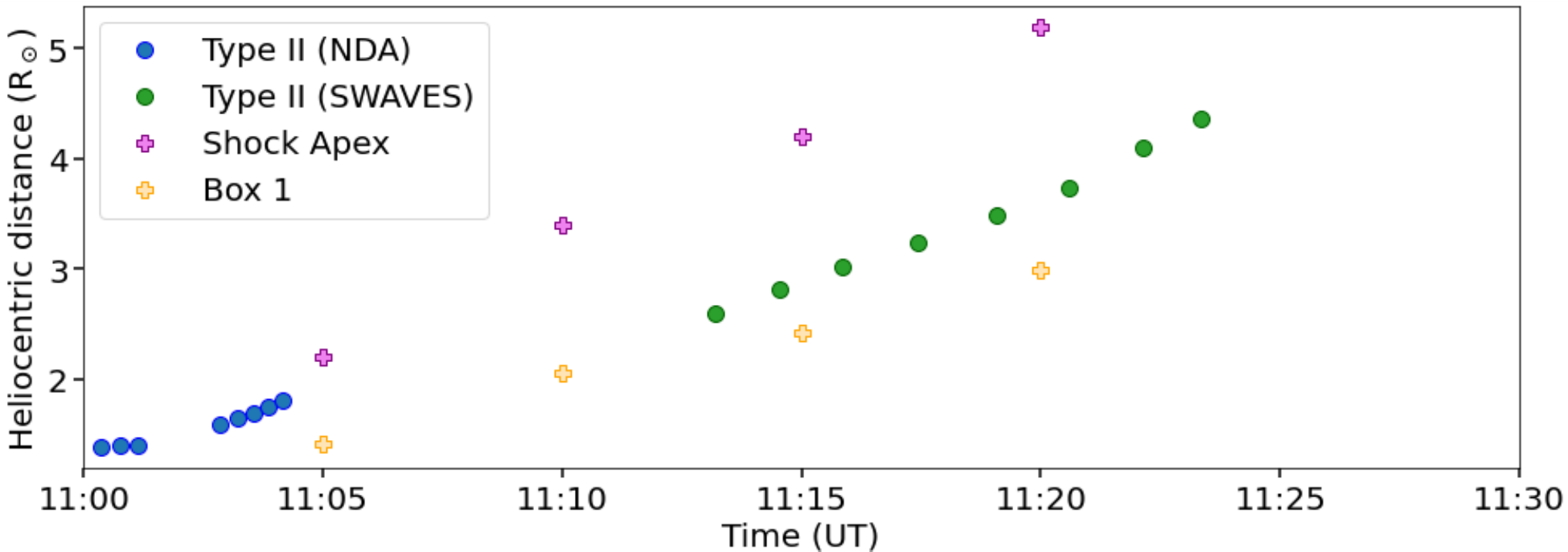}
    \caption{Heliocentric distance of two tracked type II bursts, the apex of the shock, and the location of the peak intensity in Box 1. The distances of the first (blue) and the second (green) type II bursts were calculated using their frequency-time evolution and the \cite{saito77} coronal electron density model. The distance of the apex (pink) and the point in Box 1 (yellow) were obtained from the 3D reconstruction of the shock.}
    \label{fig:apex_and_typeIIs}
\end{figure}

\section{Discussion and conclusion}

We observed herringbone-like radio emissions and type II {solar radio bursts} associated with a CME-driven shock. The herringbone-like emission exhibited a fast positive frequency drift with a drift rate of approximately $\SI{25.5}{\mega\hertz\per\second}$. Reverse drifting herringbones with such high-value drift rates have been observed before in recent observations \citep[e.g.][]{Zhang2024, morosan2024}. Additionally, the emission showed no underlying type II backbone, which has been observed before in complex radio events, such as the present study \citep[e.g.][]{morosan2022,morosan2024}. \citealt{holman1983} suggested that a quasi-perpendicular shock encountering a plasma gradient parallel to the shock front can lead to herringbone emission without the presence of a bright backbone. A recent study by \citealt{morosan2020c} suggests that the formation of magnetic traps between the shock front and overlying magnetic field can lead to the generation of herringbone-like bursts (also without a backbone) following the escape of energised electron beams. Using a 3D reconstruction of the CME-driven shock, the source locations of {this} emission were found to be co-spatial with the flank of the expanding shock. This suggests that the emission likely originates from shock-accelerated electrons. We reconstructed the shock in 3D using multi-viewpoint EUV and white-light observations. However, different approaches to 3D modelling of shocks in the low corona have been taken as well, for example, using EUV and radio observations \citep{mancuso19}. The lateral expansion speed of the shock at the locations of the radio emission resulted in speeds in the range of $\SI{800}{}-\SI{1000}{\kilo\meter\per\second}$, which indicates a fast lateral expansion. The values are consistent with previously reported values resulting from 2D analyses of fast CMEs \citep[e.g.][]{Vourlidas2003,Rigozo2011,Cheng2012,Veronig2018,mo19a,Morosan2020b}. For example, \cite{Rigozo2011} estimated an expansion speed of $\SI{1091.3}{\kilo\meter\per\second}$ within the LASCO C3 field of view for one of the CMEs they analysed. Lower in the corona, \cite{Veronig2018} found lateral expansion speeds of up to $\sim\SI{1600}{\kilo\meter\per\second}$ in the field of view of the Solar Ultraviolet Imager \citep[SUVI;][]{Seaton2018}. This CME was also associated with type II radio bursts \citep{Gopalswamy2018}. 

The relationship between the radial speed and the lateral expansion speed of CMEs and shocks has been studied previously in the field of view of SOHO/LASCO C2 and C3 coronagraphs \cite[e.g.][]{2003DalLago,Gopalswamy2009expansion,Michalek2009,Dagnew2020}. However, this relationship has not yet been studied in the low corona below $\sim2~R_\odot$. \cite{2003DalLago} studied multiple limb CMEs to determine a relation of $V_\mathrm{rad}/V_\mathrm{exp}$=0.88. Later, \cite{Gopalswamy2009expansion} found that the relation of $V_\mathrm{rad}/V_\mathrm{exp}$ is dependent on the CME width, which can result in values of less than one or greater than one. A statistical study of multiple limb CMEs in solar cycle 23 by \cite{Michalek2009} found a relation of $V_\mathrm{rad}/V_\mathrm{exp}=1.17$. \cite{Dagnew2020} compared solar cycles 23 and 24 and determined the average relations to be $V_\mathrm{rad}/V_\mathrm{exp}=0.97$ and $V_\mathrm{rad}/V_\mathrm{exp}=0.68$, respectively. In our event, we utilised the centre line of the 3D shock model, shown in Fig. \ref{fig:radial_and_lateral_distance}, to estimate the radial speed. The height of the apex for each shock reconstruction was found along the centre line. Fitting the height with a second-degree polynomial function and calculating the derivative of the fitted function resulted in radial speeds of over $\SI{2000}{\kilo\meter\per\second}$. Comparing this to the lateral expansion speeds at the location of Box 1 at the same times resulted in ratios valued between approximately $2.1$ and $2.6$. The relations are all higher than the previously reported values obtained in the corona using measurements of the lateral width of CMEs. In the present study, there is a difference of at least a factor of approximately two in these ratios for shock in the low corona and for CMEs higher in the corona within LASCO's field of view. The previous studies mainly considered the main CME body and excluded any visible shock features that might contribute to the difference in the ratios. In the low corona, however, the CME main body is relatively close in distance to the associated shock wave and their expansion speeds are likely similar.

In addition to the herringbone-like emission, type II radio bursts were observed in the frequency range of both NDA and STEREO-B/SWAVES. The frequency drift of the bursts can be tracked in the spectrum (see Fig. \ref{fig:tracking_appendix}b in Appendix \ref{app:B}) and converted to heliocentric distances using a coronal electron density model. Figure \ref{fig:apex_and_typeIIs} shows the heliocentric distance of the first (NDA, blue) and the second (SWAVES, green) type II bursts according to the coronal electron density model by \cite{saito77} as a function of time. We assumed the tracked lane of the first type II burst to be fundamental emission and the tracked lane of the second type II to be harmonic emission. The heliocentric distances of the apex of the shock (pink) and location of the intensity peak in Box 1 (yellow) are also included. Fitting a second-degree polynomial function to the heliocentric distances of both type IIs and determining their derivatives gives estimations for their speed. The first type II burst had a maximum speed of over $\SI{2300}{\kilo\meter\per\second}$, and the speeds for the seconds type II burst were approximately in the range of $\SI{1450}{}-\SI{2500}{\kilo\meter\per\second}$.

We also find a clear difference in the speed profiles in Box 1 and Box 2. The speed at the location of the source located within Box 2, which corresponds to the northernmost radio emission, is almost constant for the time period considered. However, at the location of the source within Box 1, which corresponds to the brighter, more prominent radio emission, the speed profile shows a clear deceleration over time. The decelerating profile can be fitted with a first-degree polynomial function, which is indicated by the blue dashed line in the bottom panel of Fig \ref{fig:speed_estimate}b. The fit gives the shock at the location of the radio source within Box 1 as a deceleration of $a\approx\SI{-127}{\meter\per\second\squared}$. Based on the decelerating speed profile, it is likely that the shock expanded rapidly towards the observer prior to decelerating. The results suggest that high lateral expansion speeds were achieved already during the early stages of the eruption. Such high speeds likely contributed to generating the accompanying radio emission, which is also suggested in previous studies. For example, \cite{Mann2003} studied shock waves in the corona and interplanetary space using a theoretical model and determined that a propagating disturbance can steepen to a shock wave and accelerate particles at heliocentric distances as low as $1.2$-$2.9~R_\odot$ and then beyond $6~R_\odot$. At 11:05 UT, the radio source within Box 1 on the 3D shock model is located at a heliocentric distance of $\sim1.4\,R_\odot$, thus falling inside the interval determined by \cite{Mann2003}.

Future statistical studies of the low coronal evolution of CMEs can further constrain the CME speed parameters associated with the presence of radio emissions. However, the sample of CMEs with clear shock boundaries in EUV from multiple viewpoints is likely to be low, and even lower when associated with radio events with imaging availability. Such studies may be possible in the near future with the availability of Solar Orbiter and STEREO-A moving again in orbit ahead of the Earth, which can provide the additional viewpoints necessary to study low coronal CMEs in 3D. 

\begin{acknowledgements}
{S.N. and D.E.M. acknowledge the University of Helsinki Three-Year Grant. D.E.M acknowledges the Research Council of Finland project `RadioCME' (grant number 333859) and Academy of Finland project 'SolShocks' (grant number 354409). We thank the radio monitoring service at LESIA (Observatoire de Paris) to provide value-added data that have been used for this study. We also thank the Radio Solar Database service at LESIA / USN (Observatoire de Paris) for making the NRH and ORFEES data available. We thank the eCALLISTO network for the continuous availability of radio spectra.}

\end{acknowledgements}

\bibliographystyle{aa}
\bibliography{references}

\appendix

\section{Centroids of radio emission at 173 MHz}\label{app:A}

Figure \ref{fig:centroids_appendix} shows the location of the radio emission at $\SI{173}{\mega\hertz}$ extracted from the NRH imaging for a period of $\SI{4}{\min}$ starting at 11:00 UT. The top-right panel shows the centroids from Earth's viewpoint overlaid on the SDO/AIA $\SI{211}{\angstrom}$ running difference images. The top-left panel indicates the location of the centroid from STEREO-B's perspective overlaid on the EUVI-B $\SI{195}{\angstrom}$ running difference images. The three bottom panels show the centroids together with the 3D reconstruction of the shock from the perspective of STEREO-B (left), the solar north pole (middle), and Earth (right). The centroids of the radio emission at $\SI{173}{\mega\hertz}$ exhibit a similar behaviour in time and space as the centroids at $\SI{150}{\mega\hertz}$.

\noindent\begin{minipage}{\linewidth}
\centering
\includegraphics[width=0.9\textwidth]{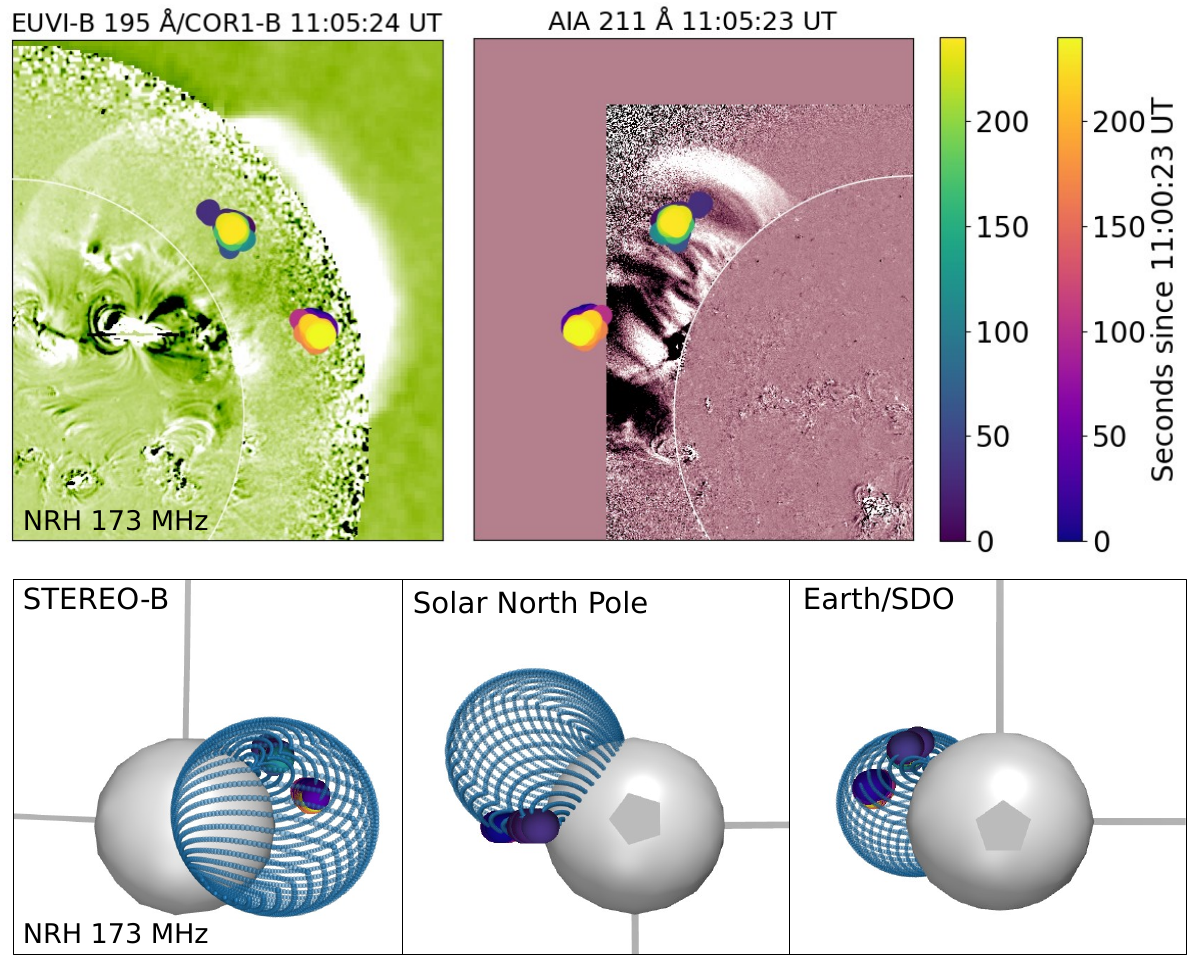}
\captionof{figure}{Centroids of the radio emission at $\SI{173}{\mega\hertz}$. The top panel shows the centroids of the radio emission from the perspective of STEREO-B (left) and Earth (right). The centroids are plotted over running difference images from EUVI at $\SI{195}{\angstrom}$ and COR1 on board STEREO-B and AIA at $\SI{211}{\angstrom}$ from SDO. The bottom panel shows the centroids together with the 3D reconstruction of the shock from three perspectives: STEREO-B (left), solar north pole (middle), and Earth (right).}
\label{fig:centroids_appendix}
\end{minipage}

\newpage

\section{Tracking the herringbone-like emission and the type II bursts}\label{app:B}

Figure \ref{fig:tracking_appendix} shows how the frequency drifts of the herringbone-like emission and the type II bursts were estimated. Figure \ref{fig:tracking_appendix}a shows a zoomed-in dynamic spectrum observed by ORFEES between $\SI{150}{\mega\hertz}$ and $\SI{190}{\mega\hertz}$ at 11:01:10-11:01:50~UT. This part of the spectrum displays the observed herringbone-like emission with reverse drifting structures. The points in the spectrum indicate the tracking of a single structure to determine its drift rate $\mathrm{d}f/\mathrm{d}t$. Figure \ref{fig:tracking_appendix}b shows dynamic spectrum observed by STEREO-B/SWAVES and NDA at $\SI{1}{}-\SI{80}{\mega\hertz}$ between 11:00 and 11:25~UT. The dots in the spectrum indicate how the type II lanes were tracked in order to infer their heliocentric distances.

\noindent\begin{minipage}{\linewidth}
\centering
\includegraphics[width=\textwidth]{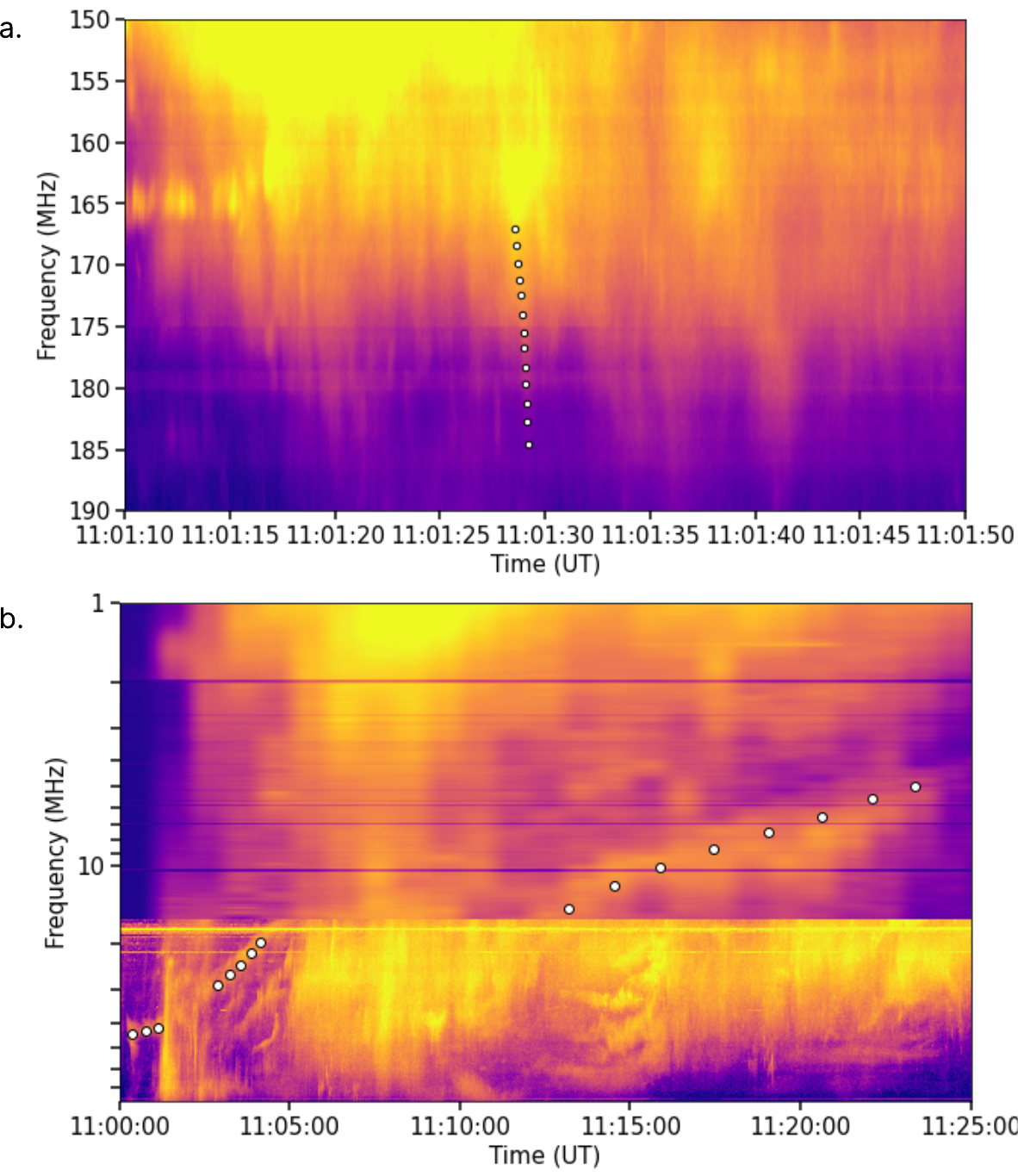}
\captionof{figure}{Tracking the herringbone-like radio emission and type II radio bursts in the dynamic spectra. (a) Dynamic spectrum observed by ORFEES at $\SI{150}{}-\SI{190}{\mega\hertz}$ between 11:01:10 and 11:01:50~UT. The points in the spectrum indicate a single reverse drifting structure and how the drift rate is determined. (b) Dynamic spectrum observed by STEREO-B/SWAVES and NDA at $\SI{1}{}-\SI{80}{\mega\hertz}$ between 11:00 and 11:25~UT.}
\label{fig:tracking_appendix}
\end{minipage}

\end{document}